\documentclass[twocolumn]{aastex61}

\usepackage{url}
\usepackage{enumitem} 
\newcommand{\red}[1]{{#1}}

\newcommand{\hMsun}{{\ifmmode{h^{-1}{\rm
        {M_{\odot}}}}\else{$h^{-1}{\rm{M_{\odot}}}$~}\fi}} 
\newcommand{\hMpc}{{\ifmmode{h^{-1}{\rm Mpc}}\else{$h^{-1}$Mpc }\fi}}
\def\be{\begin{equation}}
\def\ee{\end{equation}}
\def\ba{\begin{eqnarray}}
\def\ea{\end{eqnarray}}

\received{November 6, 2017}
\revised{November 27, 2017}
\accepted{\today}
\submitjournal{ApJ}

\shorttitle{Constraints on dynamical dark energy from AP the effect}
\shortauthors{X.-D. Li, C.G. Sabiu, C. Park, et. al}

\begin{document}

\title{Cosmological constraints from the redshift dependence of the Alcock-Paczynski effect: Dynamical dark energy}

\correspondingauthor{Cristiano G. Sabiu}
\email{csabiu@gmail.com}

\author[0000-0003-3964-0438]{Xiao-Dong~Li}
\affiliation{School of Physics and Astronomy, Sun Yat-Sen University, Guangzhou 510297, P.R.China}
\affil{School of Physics, Korea Institute for Advanced Study, 85 Heogiro, Dongdaemun-gu, Seoul 02455, Korea}

\author[0000-0002-5513-5303]{Cristiano G. Sabiu}
\affil{Department  of  Astronomy,  Yonsei  University,  50  Yonsei-ro, Seoul 03722, Korea}

\author[0000-0001-9521-6397]{Changbom~Park}
\affil{School of Physics, Korea Institute for Advanced Study, 85 Heogiro, Dongdaemun-gu, Seoul 02455, Korea}

\author{Yuting Wang}
\affil{National Astronomy Observatories, Chinese Academy of Science, Beijing, 100012, P.R.China}

\author[0000-0003-4726-6714]{Gong-bo Zhao}
\affil{National Astronomy Observatories, Chinese Academy of Science, Beijing, 100012, P.R.China}

\author[0000-0002-7464-7857]{Hyunbae Park}
\affil{Korea Astronomy and Space Science Institute, 776, Daedeokdae-ro, Yuseong-gu, Daejeon, 34055, Korea}

\author[0000-0001-6815-0337]{Arman Shafieloo}
\affil{Korea Astronomy and Space Science Institute, 776, Daedeokdae-ro, Yuseong-gu, Daejeon, 34055, Korea}
\affil{University of  Science and Technology (UST), Yuseong-gu 217 Gajeong-ro, Daejeon 34113, Korea}

\author[0000-0002-4391-2275]{Juhan Kim}
\affil{Center for Advanced Computation, Korea Institute for Advanced Study, 85 Hoegi-ro, Dongdaemun-gu, Seoul 02455, Korea}

\author[0000-0003-4923-8485]{Sungwook E. Hong}
\affil{Korea Astronomy and Space Science Institute, 776, Daedeokdae-ro, Yuseong-gu, Daejeon, 34055, Korea}



\begin{abstract}

We perform an anisotropic clustering analysis of 1,133,326 galaxies from the Sloan Digital Sky Survey (SDSS-III) Baryon Oscillation Spectroscopic Survey (BOSS) Data Release (DR) 12 covering the redshift range $0.15<z<0.69$.
The geometrical distortions of the galaxy positions, caused by incorrect cosmological model assumptions, are captured in the anisotropic two-point correlation function on scales  6 -- 40 $h^{-1}\rm Mpc$. The redshift evolution of this anisotropic clustering is used to place constraints on the cosmological parameters. We improve the methodology of Li et al. 2016, to enable efficient exploration of high dimensional cosmological parameter spaces, and apply it to the Chevallier-Polarski-Linder parametrization of dark energy, $w=w_0+w_a{z}/({1+z})$.
In combination with the CMB, BAO, SNIa and $H_0$ from Cepheid data, we obtain
$\Omega_m = 0.301 \pm 0.008,\ w_0 = -1.042 \pm 0.067,\ $ and $w_a = -0.07 \pm 0.29$ (68.3\% CL). 
Adding our new AP measurements to the aforementioned results 
reduces the error bars by $\sim$30 -- 40\% and improves the dark energy figure of merit by a factor of $\sim$2. 
We check the robustness of the results using realistic mock galaxy catalogues. 

\end{abstract}




\section{Introduction}

\begin{figure*}
   \centering{
   \includegraphics[height=5cm]{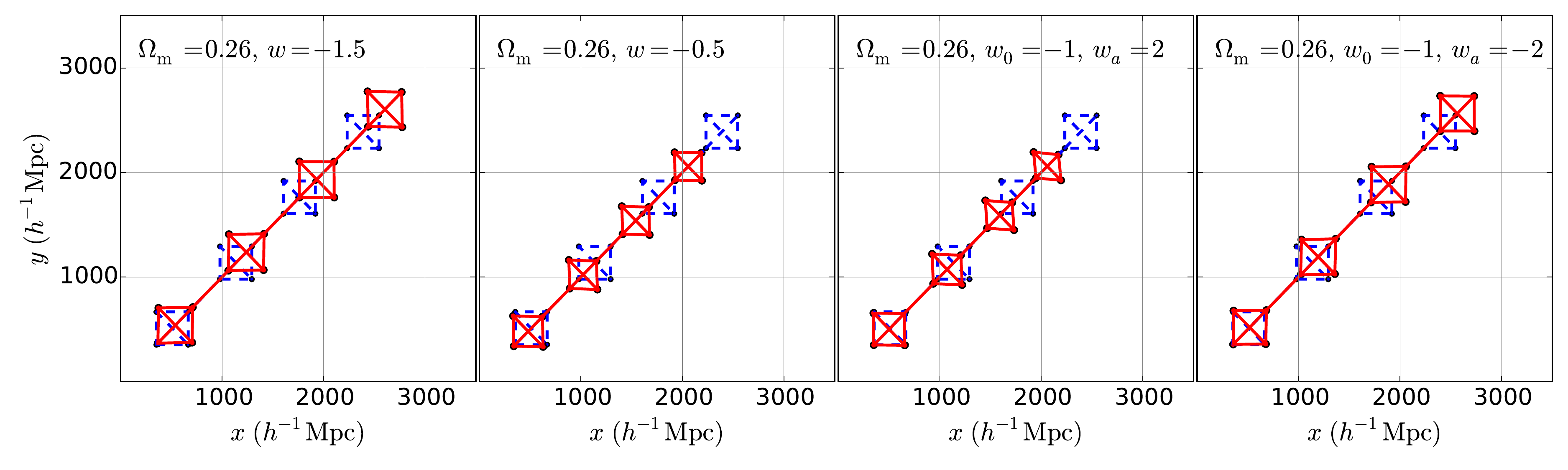}
   }
   \caption{\label{fig_xy}
   Examples of the rectangular shape distorted by assuming incorrect 
   cosmologies compared to the true fiducial cosmology $\Omega_m=0.26$ and $w=-1$. \red{In 2D comoving coordinates with the observer at the origin, 4 perfect squares are plotted at various distances along one particular line-of-sight direction, in the fiducial model (blue).  These squares are then reprojected into am incorrect cosmological model (red), distorting only the radial positions of the corners of each square and resulting in a distorted quadrilateral shapes.}
   }
\end{figure*}


\begin{figure}
   \centering{
   \includegraphics[width=\columnwidth,trim=0 1cm 31cm 0, clip]{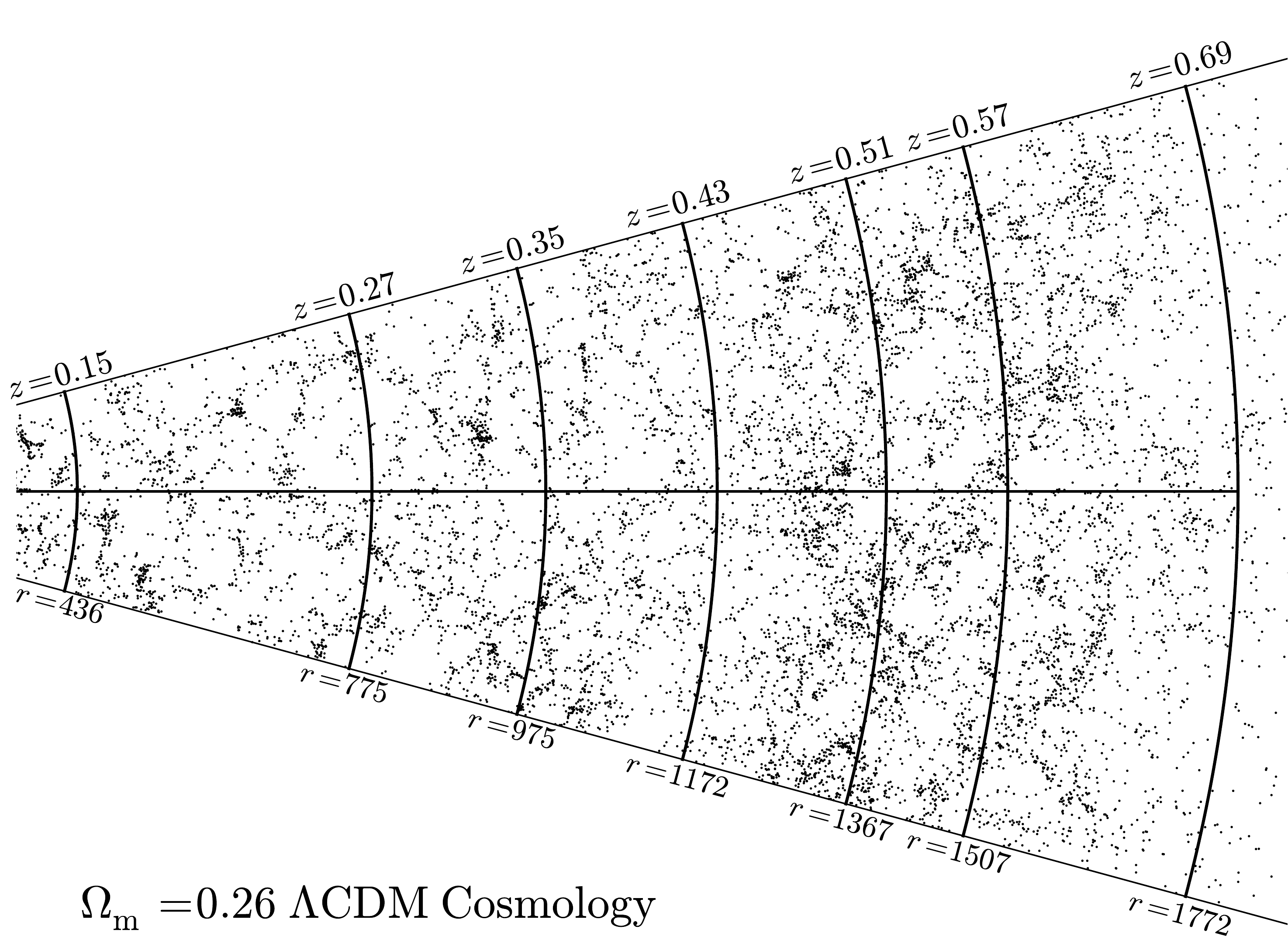}
   }
   \caption{\label{fig_fan}
   A patch of SDSS BOSS DR12 galaxies in the fan shape region of 
   $\rm 140^\circ < R.A. < 170 ^\circ$, $\rm 10^\circ < Decl. < 13 ^\circ$,
   split into six non-overlapping redshift bins (marked by the arcs) in order to probe the redshift evolution of anisotropic clustering.
   We plot the galaxy positions computed using $\Omega_m$=0.26 $\Lambda$CDM.
   Redshifts and comoving distances (in unit of $h^{-1}\rm Mpc$) of the edges of redshift bins are listed.
   %
   }
\end{figure}

The  origin  of  the late-time accelerating  expansion  of  the  universe  is  one of  the  most  salient  questions  in  contemporary cosmology. 
Theoretical explanations for this phenomena are numerous and range from a non-zero  vacuum energy, an evolving  scalar  field  remnant  from  the  big  bang, 
to modifications of Einstein's General Relativity \citep{Li2011,2012IJMPD..2130002Y,2015PhR...568....1J}. 
Considering the wealth of theoretical explanations, 
it is crucial to obtain precise and unbiased measurements of the expansion history of the Universe which allows us to 
differentiate between competing models. 

In recent years the Alcock-Paczynski (AP) test \citep{AP1979} applied to galaxy redshift samples \citep{Outram2004,Blake2011,Alam2016}, 
has allowed tight constraints to be placed on the background averaged distance scales, $D_A(z)$ and $H^{-1}(z)$.  
Assuming an incorrect cosmological model for the coordinate transformation between redshift space and comoving space
produces residual geometric distortions in the resultant galaxy distribution as well as a change in volume elements \citep{topology}, see Figure \ref{fig_xy} as an illustrative example. 
These distortions are induced by the fact that measured distances along 
and perpendicular to the line of sight depend on the given cosmological parameters. 
Therefore, measuring the ratio of galaxy clustering in the radial and transverse directions provides a probe of this AP effect, which is sensitive to the product $D_A(z)H(z)$.  


The main caveat in applying the AP test is that 
the radial distances of galaxies are inferred from observed redshifts.
Thus AP tests are inevitably affected by the peculiar motions of galaxies,
which leads to apparent anisotropy in the clustering signal, even if the adopted cosmology is correct.
The effect, known as redshift-space distortions (RSD),
is notoriously difficult to model accurately in the statistics of galaxy clustering \citep{Ballinger1996}.

The symmetry properties of galaxy pairs \citep{Marinoni2010,BB2012}  could also be used to probe the AP effect;
however, since the peculiar velocity distorts the redshifts and changes the apparent tilt angles of galaxy pairs,
this method is also seriously limited by RSD \citep{Jennings2011}.

In an effort to minimize RSD contamination, the shape of void regions \citep{Ryden1995,LavausWandelt1995,2016PhRvL.117i1302H}  has been 
proposed as an AP probe. This approach has the advantage that the void regions are easier to model compared with dense regions, 
but has limitations in that it utilizes only low density regions of the LSS and requires large samples to attain statistical significances and achieve competitive constraints \citep{Qingqing2016}.


Previously, we proposed to use the {\it redshift dependence} of the AP distortion \citep{Li2014} as a way of mitigating the RSD effect. 
The clustering anisotropies produced by RSD are, although large, close to uniform in magnitude over a wide range in redshift.  
However, if cosmological parameters are incorrectly chosen and there exists the AP effect, 
the anisotropy in the clustering signal has a clear redshift dependence
(as an illustration, Figure \ref{fig_xy} shows how the shape distortion varies with distance 
when incorrect cosmologies are used to infer distance from redshift).
In \cite{Li2015}, we developed an AP methodology 
that utilizes the redshift dependence of the galaxy 2-point correlation function (2PCF), 
measured as a function of angle between the galaxy pair and line-of-sight (LoS).

\begin{figure*}
   \centering{
   \includegraphics[width=1.55\columnwidth]{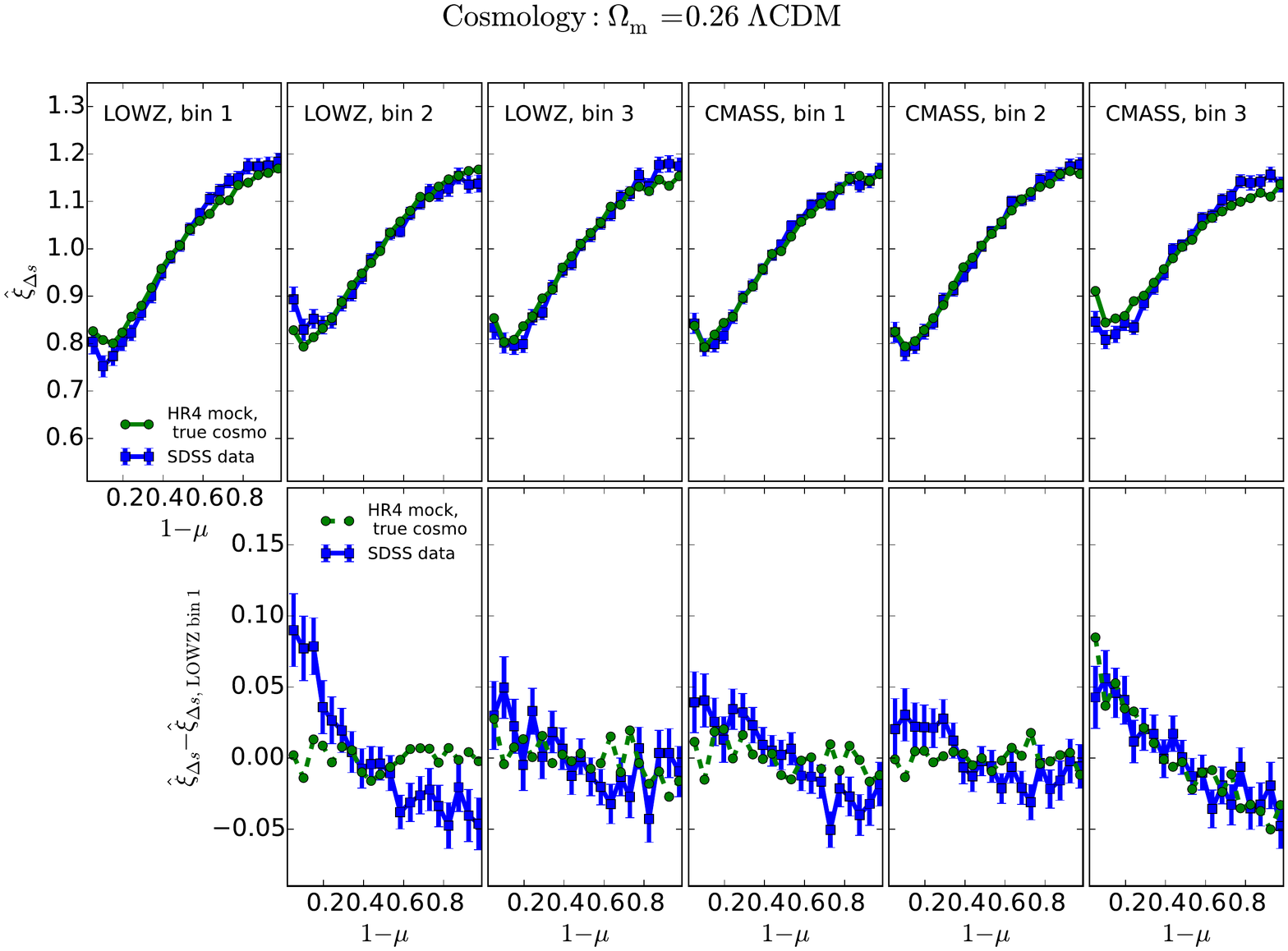}
   \includegraphics[width=1.55\columnwidth]{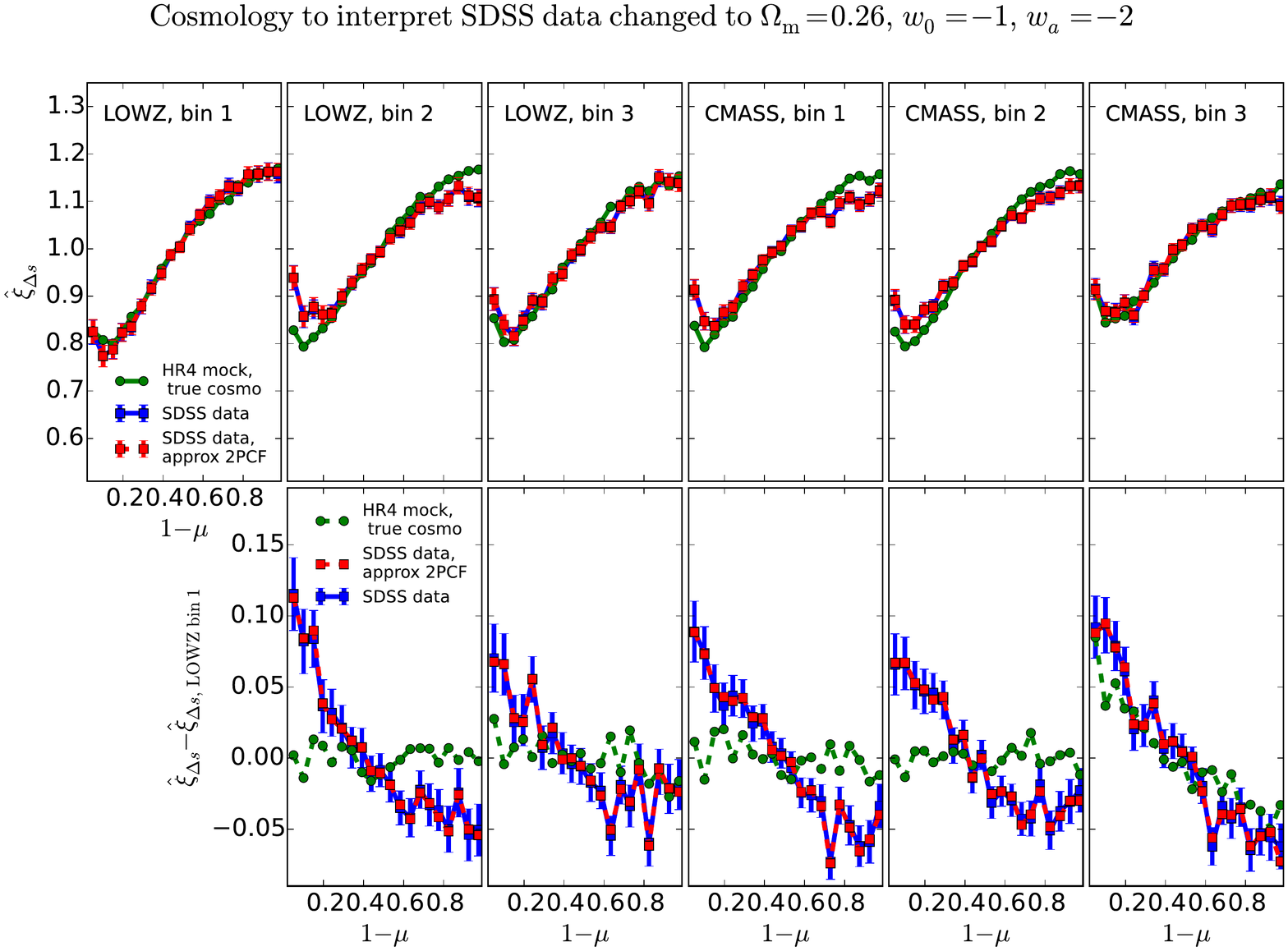}
   }
   \caption{\label{fig_xi}
   $\hat\xi_{\Delta s}(\mu)$ measured from the SDSS BOSS DR12 galaxies in six redshift bins (three in LOWZ and three in CMASS),
   assuming the $\Omega_m=0.26$ $\Lambda$CDM cosmology and a more dark energy dominated cosmology with $w_a=-2$.
   \red{Measurements, without systematic correction, are plotted for each of the six redshift bins and their redshift evolution with respective to the first bin of LOWZ.}
   In the $w_a=-2$  cosmology, the shapes of $\hat\xi_{\Delta s}(\mu)$ are different from the $\Omega_m=0.26$ cosmology results,
   and the difference changes with redshift;
   a large redshift evolution of $\hat\xi_{\Delta s}(\mu)$ is detected in this cosmology, 
   indicating that it is not likely to be the underlying true cosmology of our universe.
   The measurements in the HR4 mock catalogues (always in the $\Omega_m=0.26$ $\Lambda$CDM cosmology; plotted in green color)
   match the general shape of curves measured from observational data, 
   indicating that the simulation  reproduces the FoG and Kaiser effects.
   For the $w_a=-2$ cosmology, we also plot the approximate 2PCFs (red dashed lines) 
   inferred using the technique described in Appendix. \ref{sec:approx_2pcf}.
   The error induced in the approximation procedure is very small.
   }
\end{figure*}

\red{In an earlier work (Li et al. 2016, hereafter L16) we applied this AP method to galaxies from the Sloan Digital Sky Survey (SDSS-III), data release (DR) 12.}
Combining the method with measurements of the Cosmic Microwave Background (CMB), type Ia supernovae (SNIa), 
baryon acoustic oscillations (BAO), and $H_0$,
we obtained very tight constraints of $ \Omega_m = 0.301 \pm 0.006,\ w=-1.054 \pm 0.025$.
In reducing the RSD effect, 
we were able to use galaxy clustering on scales down to 6 $h^{-1}$Mpc,
which is a major advance in extracting cosmological information 
on small scales where galaxy clustering is strong and many independent structures exist.

In this paper, we continue to develop our previous methodology and proceed to set the constraints on dynamical dark energy. 
\red{We will use the same observational data as in L16, however we present an improved methodology compared to L16, 
allowing for faster likelihood estimation and thus the exploration of larger, higher dimensional parameter spaces.} 
The methodology we will present here, can be applied to any model of dynamical dark energy, 
or indeed any appropriately chosen parametric or non-parametric decomposition of the cosmic expansion history. 
However, as a first step, in this paper we will focus on the widely used Chevallier-Polarski-Linder (CPL) parametrization\citep{CPL_CP,CPL_L},
\begin{equation}
w(z) = w_0 + w_a (1-a) = w_0 + w_a \frac{z}{1+z}.
\end{equation}
This parametrization characterizes the dark energy equation-of-state (EoS) by two free parameters;
$w_0$ determines the present-day value, while parameter $w_a$ characterizes the first-order derivative of $w$ with respect to $a$. 
The possible redshift evolution of dark energy EoS is not considered in the analysis of L16. 

The CPL parametrization has many obvious advantages, for instance, a manageable parameter space, 
the bounded behavior at high redshift, 
and the ability to accurately reconstruct many dark energy theories \citep{CPL_L}. 
The constraining power is usually quantified by the DETF  \citep[Dark Energy Task Force;][]{DETF} figure of merit, 
defined as the reciprocal of the area of the error ellipse enclosing the 95\% confidence limit (CL) in the $w_0 - w_a$ plane. 

The rest of this paper is arranged as follows; in \S\ref{sec:method} we describe the data that is used and outline the methodology for obtaining cosmological constraints. In \S\ref{sec:results} we present the main findings of this study and finally we conclude in \S\ref{sec:conclusion}.

\section{Methodology}\label{sec:method}

\red{The methodology follows closely that of our previous work, Li et al. 2016, where we used the redshift dependence of the anisotropic clustering of galaxies to test cosmological models. When transforming galaxy positions in \{RA,DEC, Redshift\} to comoving cartesian coordinates we must assume a cosmological model. Any difference between our assumed model and the true model will induce geometrical distortions on the resultant galaxy distribution (AP effect). This can be more easily visualised in Figure \ref{fig_xy}, where we  illustrate the AP effect in four incorrect cosmologies. 

In this toy model, the boxes in the fiducial cosmology (blue) are reprojected into different cosmologies (red) with various choices of $\Omega_m, w_0, w_a$. As we can see from the figure, varying the cosmology alters the position, size and shape of the boxes in a redshift dependent fashion. Thus, we may expect that the shape of the clustering statistics will also be effected by in a similar way.}

In Li et al. 2016 we considered only non-evolving DE models. However, since 
a redshift dependence of the shape distortion is observed when adopting wrong values of $w_0$ and $w_a$, in Figure \ref{fig_xy}, we expect that these two parameters will be sensitive to our method.

\subsection{Data}

We use the spectroscopic galaxy sample of SDSS-III BOSS (Baryon Oscillation Spectroscopic Survey), which has two primary catalogues:
the LOWZ sample, designed as an extension of the SDSS-I/II luminous red galaxy sample to $z\approx 0.4$ and fainter luminosities,
and the CMASS sample covering a higher range ($0.4\lesssim z \lesssim 0.7$),
and made to be an approximately stellar mass limited sample of massive, luminous galaxies \citep{Reidetal:2016}.
In the clustering analysis we use 1,133,326 galaxies, split into six, non-overlapping redshift bins, 
$0.150<z_1<0.274<z_2<0.351<z_3<0.430<z_4<0.511<z_5<0.572<z_6<0.693$.
The edges are determined so that the number of galaxies are roughly the same in different redshift bins 
(for LOWZ and CMASS samples, respectively).

Figure \ref{fig_fan} shows a patch of 10,976 BOSS DR12 galaxies,
whose positions are computed in the $\Omega_m$=0.26 $\Lambda$CDM cosmology. 
By investigating how the anisotropy of galaxy distribution evolves in the six redshift bins,
we are able to distinguish particular cosmological models.

\subsection{Quantifying the redshift dependence of the AP distortion}


Following our previous methodology, the information of anisotropic clustering is computed\footnote{These correlations were computed using the public code \texttt{KSTAT} \url{https://bitbucket.org/csabiu/kstat}.}  as 
\begin{equation}
\xi_{\Delta s} (\mu) \equiv \int_{s_{\rm min}}^{s_{\rm max}} \xi (s,\mu)\ ds,
\end{equation}
with $s_{\rm min}=6 h^{-1} {\rm Mpc},$ and $s_{\rm max}=40 h^{-1} {\rm Mpc}$.
We then normalize these {\em clustering shells} as 
\begin{equation}
\hat\xi_{\Delta s}(\mu) \equiv \frac{\xi_{\Delta s}(\mu)}{\int_{0}^{\mu_{\rm max}}\xi_{\Delta s}(\mu)\ d\mu}
\end{equation}
to nullify the amplitude information of the clustering signal, 
which is not associated with the AP test and is mostly sensitive to the galaxy bias evolution.
The ``correct'' cosmological model is selected by minimizing the amount of redshift evolution of $\hat\xi_{\Delta s}$,
via a $\chi^2$ function of 
\begin{equation}
 \chi^2\equiv \sum_{i=2}^{6} \sum_{j_1=1}^{n_{\mu}} \sum_{j_2=1}^{n_{\mu}} {\bf p}(z_i,\mu_{j_1}) ({\bf Cov}_{i}^{-1})_{j_1,j_2}  {\bf p}(z_i,\mu_{j_2})
\end{equation}
where ${\bf p}(z_i,\mu_{j})$ is the redshift evolution of clustering with respect to the lowest redshift bin,
while subtracting systematic effects as
\begin{eqnarray}
 {\bf p}(z_i,\mu_{j}) \equiv\ & \left[\hat\xi_{\Delta s}(z_i,\mu_j)-\hat\xi_{\Delta s}(z_1,\mu_j)\right] \\ \nonumber
 &- \left[\hat\xi_{\Delta s}(z_i,\mu_j)-\hat\xi_{\Delta s}(z_1,\mu_j)\right]_{\rm sys}.
\end{eqnarray}
We use $n_{\mu}$=20, 21, ... 25 bins for the value of 
$0<\mu<\mu_{\rm max}$.
To reduce the fiber collision and the finger of god (FoG) effect \citep{FOG} near the LOS we take a cut $\mu_{\rm max} = 0.97$.

\begin{figure*}
   \centering{
   \includegraphics[width=0.9\columnwidth]{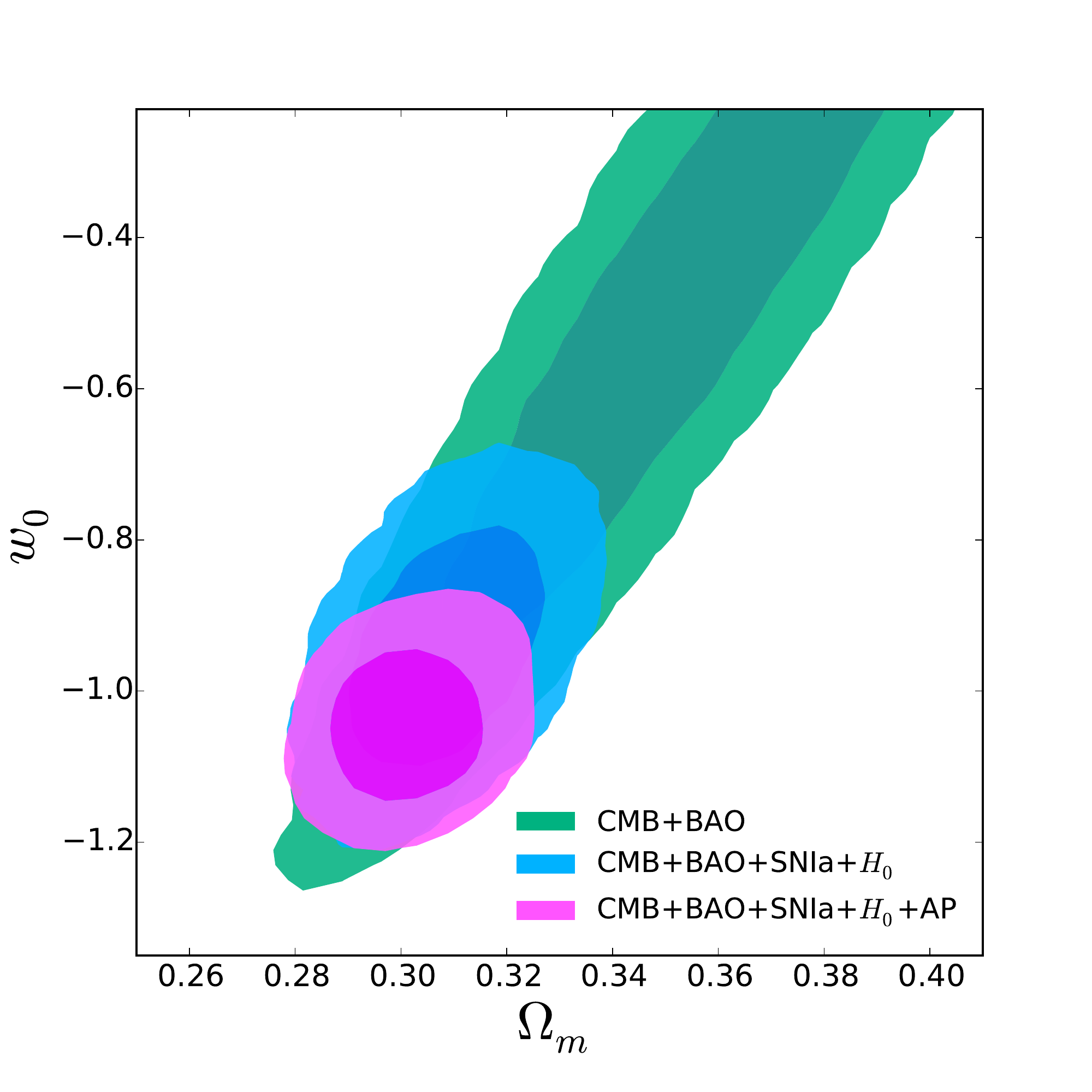}
   \includegraphics[width=0.9\columnwidth]{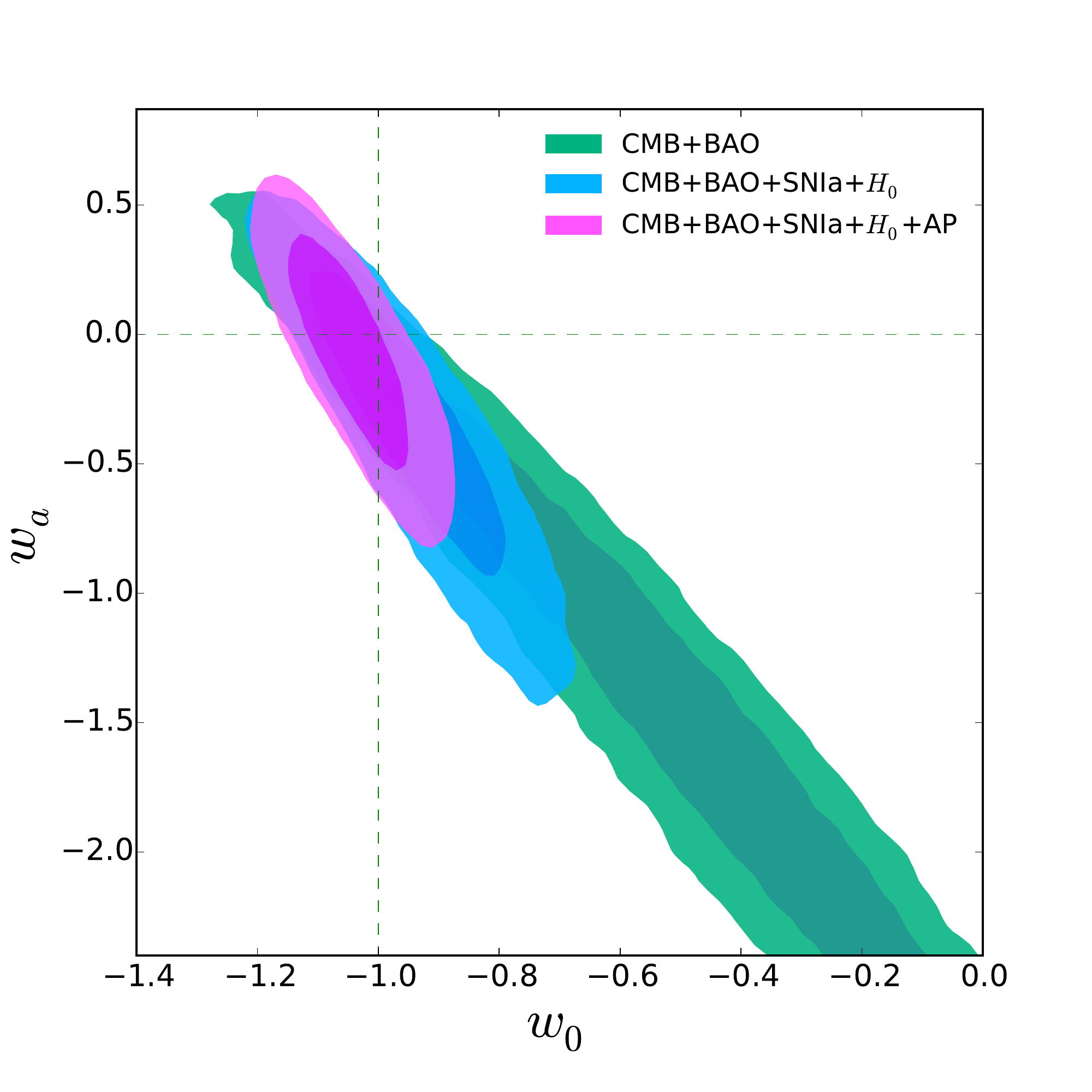}
   }
   \caption{\label{fig_con}
   Cosmological parameter constraints on the CPL dark energy parametrization $w=w_0+w_a {z}/{(1+z)}$.
   The 68.3\%, 95.4\% CL likelihood contours in the $\Omega_m - w_0$  and $w_0 - w_a$ planes are plotted in the left and right panels, respectively.
   Results from CMB+BAO (cyan filled), CMB+BAO+SNIa+$H_0$ (blue filled)
   and CMB+BAO+SNIa+$H_0$+AP (magenta filled) are shown.
   Adding our AP method to the CMB+BAO+SNIa+$H_0$ combination reduces the contour area by as much as 50\%.  
   }
\end{figure*}

The systematics effects are estimated using mock catalogues drawn from Horizon Run 4 \citep[HR4;][]{HR4},
an N-body simulation with a box size of $L={3150}$ $h^{-1}$Mpc, the number of particles $6300^3$,   
initial redshift of $z_{i}=100$, and the WMAP5\citep{komatsu2011} cosmological parameters 
$(\Omega_{b},\Omega_{m},\Omega_\Lambda,h,\sigma_8,n_s)$  = (0.044, 0.26, 0.74, 0.72, 0.79, 0.96). 
Mock galaxy samples are produced using a modified version of the one-to-one correspondence scheme \citep{hong2016}. 

\red{Since the mock catalogues were analysed using the cosmology with which they were run, they have no geometrical distortions associated with the AP effect, allowing us to focus solely on modeling the RSD effect.}

The covariance matrix, ${\bf Cov}$, is computed from the a set of 2,000 MultiDark PATCHY mock catalogues \citep{MDPATCHY}.
The statistical bias and scattering in the likelihood function (due to the finite number of mocks in covariance estimation) 
are adequately corrected \citep{Hartlap,Percival2014}.

The MultiDark PATCHY  mocks are produced using approximate gravity solvers and analytical-statistical biasing models.
They were calibrated to the BigMultiDark N-body simulation \citep{K2014}, which 
uses $3\,840^3$ particles in a volume of $(2.5h^{-1}\rm Gpc)^3$,
assuming a $\Lambda$CDM cosmology with 
$(\Omega_{b},\Omega_{m},h,\sigma_8,n_s)$  = (0.048206, 0.307115, 0.6777, 0.8288, 0.9611). 
The mock surveys can well reproduce the number density, 
selection function, survey geometry, and 2PCF measurement of the BOSS DR12 catalogues.
They have been adopted for statistical analysis of BOSS data in a series of works \citep[see][and references therein]{Alam2016}.

As an illustration,
Figure \ref{fig_xi} shows how we us the above procedure to distinguish different cosmologies.
Here we plot the value of $\hat\xi_{\Delta s}(\mu)$ (upper panels) as well as its redshift evolution (lower panels), 
measured from the BOSS DR12 galaxies in six redshift bins.
Two cosmologies are adopted, 
one with $\Omega_m=0.26$, and the other with a strongly disfavoured value of $w_a=-2$.

The shape of $\hat \xi_{\Delta s}(\mu)$ is very different from a flat curve, 
due to the apparent anisotropy produced by the peculiar motion of galaxies.
In the $w_a=-2$ cosmology, 
the shapes of $\hat\xi_{\Delta s}(\mu)$ are different from the measurements in the $\Omega_m=0.26$ cosmology,
and the amount of difference systematically evolves with redshift.
We observe a large redshift evolution of $\hat\xi_{\Delta s}(\mu)$, 
indicating that it is not likely to be the underlying true cosmology of our universe.

The green curves denote the 2PCFS measured from the HR4 mock catalogues
(we plot the {\it correct} measurement in the simulation cosmology, i.e. the $\Omega_m=0.26$ $\Lambda$CDM) \red{and have not been corrected for systematics. So their amplitude simply represents the magnitude of the systematic effects.}

The simulation results can match the general shape of the results from observational data,
indicating that the FoG \citep{FOG} and Kaiser \citep{Kaiser1987} effects are both well reproduced.
Since there is no AP effect in the simulation measurements,
all detected redshift evolution should be due to effects other than the cosmological effect; 
so they are adopted as an estimation of the systematic effects of the method.
The amount of systematics reaches 4 -- 6\% in the 6th redshift bin,
and is much smaller ($\lesssim2\%$) in the other bins.

In L16, the likelihood contour of $\Omega_m - w$ was constructed by
measuring the 2PCF 3,375 times,
using 3D positions of BOSS galaxies computed in 71$\times$45 sets of cosmological parameters.
This procedure took $\sim$1 month using 500 cores of the Korea Institute for Advanced Study {\texttt {Baekdu}} cluster.
It would be computationally intractable to attempt a full MCMC of all relevant cosmological parameters using this approach. 
Thus we adopt an ``approximate 2PCF'' by transforming our measurements from one cosmology to another. A detailed explanation of this procedure is given in Appendix \ref{sec:approx_2pcf}

\section{Cosmological constraints}\label{sec:results}

The Planck team has released the {\texttt {COSMOMC}} \citep{LB2002} outputs of four Markov Chain Monte Carlo (MCMC) 
``chains'' in the CPL model, 
using a combination of four datasets:
the full-mission Planck observations of CMB temperature and polarization anisotropies \citep{Planck2015},
the BAO distance priors measured from SDSS DR11 \citep{Anderson2013}, 6dFGS \citep{6dFGS} and SDSS MGS \citep{MGS},
the ``JLA'' SNIa sample \citep{JLA},
and the Hubble Space Telescope measurement of $H_0=70.6\pm3.3$ km/s/Mpc \citep{Riess2011,E14H0}.
These MCMC chains contain the CMB+BAO+SNIa+$H_0$ likelihood computed for $\sim$37,000 sets of cosmological parameters.
After adding the log-likelihoods of the Planck team sample with ours, 
while also multiply the sample weights by our likelihoods, 
we derive the CMB+BAO+SNIa+$H_0$+AP constraints on CPL parameters. 

\subsection{Results}


Figure \ref{fig_con} shows the 68.3\% and 95.4\% CL likelihood contours in the $\Omega_m - w_0$  and $w_0 - w_a$ planes,
derived from the CMB+BAO, CMB+BAO+SNIa+$H_0$ and CMB+BAO+SNIa+$H_0$+AP, respectively.
The overlapping of the various contours suggest that they are consistent with each other.



The current CMB+BAO datasets are not statistically powerful enough to effectively constrain the w0-wa parameter space.
Combining the four external techniques, i.e. CMB+BAO+SNIa+$H_0$, leads to effective constraints on all parameters.
The statistical mean values and 68.3\% uncertainties of these parameters are
\begin{eqnarray}
&\Omega_m = 0.309 \pm 0.010,\\
&w_0 = -0.938 \pm 0.109,\\
&w_a = -0.38 \pm 0.41,
\end{eqnarray}
while adding our AP method to this combination further tightens the constraints leading to
\begin{eqnarray}
&\Omega_m = 0.301 \pm 0.008,\\
&w_0 = -1.042 \pm 0.067,\\
&w_a = -0.07 \pm 0.29.
\end{eqnarray}
The error bars are dramatically reduced by 30 -- 40\%,
and the contour areas are reduced by $\sim$50\%, i.e., the dark energy figure of merit is improved by $\sim$100\%. 
Notice that the AP constraints come from the BOSS DR12 data, which is already used in the BAO analysis.
So the doubling of the figure of merit comes at no additional cost or alteration to data size,
thus greatly improves the overall cost-benefit of the large cosmological redshift surveys.

\citet{2018arXiv180107403Z} tested the correlation between the BAO and AP methods, and find that the information extracted from each methods is statistically independent. The BAO method uses the BAO feature in the clustering of galaxies on scales of 100-150 $h^{-1}$Mpc, created by the oscillation of the baryon-photon plasma in the early Universe. Measuring the BAO feature in 1D or 2D then yields measurements of $D_V$ or $D_A$ and $H$ at some representative redshift. As a comparison, the AP method uses galaxy clustering on scales of $6--40\ h^{-1}$Mpc, which is much smaller than the BAO scale. The information explored from the two methods are fairly independent, so we can easily combine them without worrying about their correlation.

\begin{figure}
   \centering{
   \includegraphics[width=\columnwidth]{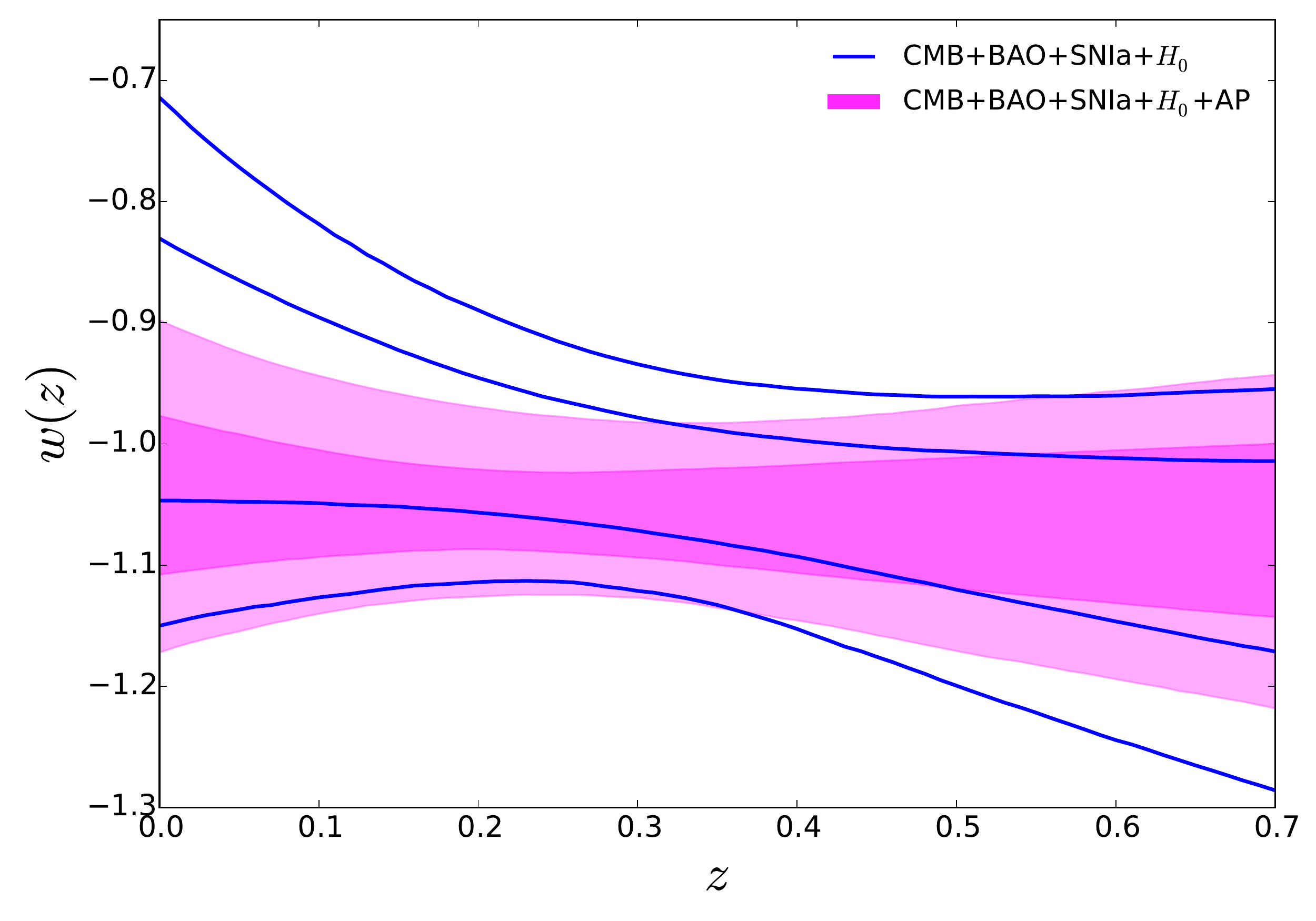}
   }
   \caption{\label{fig_wz}
   Derived redshift evolution of $w(z)$, the 68.3\% and 95.4\% CL regions are plotted.
   Adding the AP method tightens the constraints and reduces the redshift evolution of $w$ (the tilt of $w(z)$).
   }
\end{figure}

It can be also noted that, after adding the AP method, the central value of $w_a$ moved significantly towards zero.
This means the result becomes more consistent with a cosmological constant dark energy component having no evolution.
Figure \ref{fig_wz} shows the redshift evolution of $w(z)$ derived from the cosmological constraints.
Adding the new AP results tightens the constraints and reduces the redshift evolution of $w$ (tilt of $w(z)$).

In Appendix \ref{sec:robust}, these results are tested for robustness. We find that the results are unaffected by the line-of-sight $\mu$-cut, 
the range of radial integration, the choice of fiducial cosmology in the mapping of $\xi(s,\mu)$,
and the number of mocks.

\section{Conclusions}\label{sec:conclusion}

In recent studies we have proposed to constrain cosmological parameters 
governing the expansion history of the universe via 
the redshift dependence of anisotropic galaxy clustering \citep{Li2014,Li2015,Li2016}.
This approach enables a robust AP test on relatively small scales.
In this paper we improved the methodology and obtained constraints 
on the CPL parametrization of dark energy.
The derived cosmological constraints are fully consistent with a cosmological constant 
dark energy component having no redshift evolution.


The AP method presented in this work has many advantages over the 
`traditional' methods using galaxy clustering.
Since it works with the redshift evolution of the anisotropic clustering signal it significantly reduces the effect of systematics. 
Our method mitigates many of the difficulties in accurately  modeling the RSD, non-linear clustering and galaxy bias.
This implementation of the AP test can use galaxy clustering statistics at relatively smaller scales compared to other methods, thus including many more k-modes, maximising the information gain.

In this analysis, we find that the systematic effects do not significantly affect the derived cosmological constraints.
But it remains to be seen if this is true for future galaxy surveys.
In particular, the systematic effects are estimated using simulations performed in one fiducial cosmology.
The cosmological dependence of the systematics remains to be investigated in future works.

In this analysis, combining our method with the CMB+SNIa+BAO+$H_0$ datasets,
the dark energy figure of merit is improved by a factor of $\sim$2.
This indicates the great power of the method in constraining the cosmic expansion history and probing the properties of dark energy.

\red{In \citet{Li2014,Li2015} we tested and found that our methodology is applicable up to z~1.5. Thus future surveys such as EUCLID and DESI will provide ideal data for the method presented in our current and previous works.}

\red{Previously we found that, for a $1/8$-sky mock surveys having 8 million galaxies and sampled to have roughly a uniform number density in $z=0-1.5$, the AP effect results in tight constraints with 68.3\% CL intervals of $\delta \Omega_m\sim0.03$ and $\delta w\sim0.1$ when using the AP test alone, without combining with others. The constraints from DESI, which will probe 30 million galaxies and reach $z=1.5$, will be tighter than that.}

It would be interesting to see whether we can detect firm evidence for w(z) deviation from -1 in future surveys. This also demands us make more precise correction of systematics, which would becomes comparable to or even larger than the  statistical error.

\red{Although at the level of precision of current surveys the cosmological dependence of the systematic is negligible as can be seen in appendix B. In the era of next stage experiments, the impact of cosmological dependence of systematics would be definitely larger. If we assume the statistical error proportional to $1/\sqrt{N}$ where $N$ being the number of galaxies, then future surveys such as DESI will have $\sim$6 times smaller statistical error than SDSS-III. However, the cosmological dependence of the systematic  could be easily solved by, e.g. interpolating among systematics estimated from several sets of simulations with different cosmologies, or considering theoretical estimation of systematics \citep{park18}. Thus we believe this would not be a significant problem limiting the application of the method.}

We expect the method will play an important role in deriving cosmological constraints from future spectroscopic galaxy surveys.

\acknowledgments

We thank Korea Institute for Advanced Study for providing computing resources (KIAS Center for Advanced Computation Linux Cluster System).
CGS acknowledges support from the National Research Foundation (NRF-2017R1D1A1B03034900). 
A.S. would like to acknowledge the support of the National Research Foundation of Korea (NRF-2016R1C1B2016478).

We thank Stephen Appleby, Seokcheon Lee, Maurice van Putten, Graziano Rossi and Yi Wang for helpful discussions.

Based on observations obtained with Planck (\url{http://www.esa.int/Planck}), 
an ESA science mission with instruments and contributions directly funded by 
ESA Member States, NASA, and Canada.

Funding for SDSS-III has been provided by the Alfred P. Sloan Foundation, the Participating Institutions, the
National Science Foundation, and the U.S. Department of Energy Office of Science. 
The SDSS-III web site is \url{http://www.sdss3.org}. 
SDSS-III is managed by the Astrophysical Research Consortium for the Participating Institutions
of the SDSS-III Collaboration including the University of Arizona, the Brazilian Participation Group, Brookhaven
National Laboratory, Carnegie Mellon University, University of Florida, the French Participation Group, 
the German Participation Group, Harvard University, the Instituto de Astrofisica de Canarias, the Michigan State/Notre
Dame/JINA Participation Group, Johns Hopkins University, Lawrence Berkeley National Laboratory, Max Planck
Institute for Astrophysics, Max Planck Institute for Extraterrestrial Physics, New Mexico State University, New
York University, Ohio State University, Pennsylvania State
University, University of Portsmouth, Princeton University,
the Spanish Participation Group, University of Tokyo, University of Utah, Vanderbilt University, University of Virginia, 
University of Washington, and Yale University.

\appendix

\section{Approximating the 2PCFs in cosmologies other than the fiducial one}\label{sec:approx_2pcf}

\red{The number of galaxy pairs are counted in bins of separation, $s$ and cosine of the angle w.r.t the LoS $\mu$, where the comoving positions were computed in a fiducial cosmology ($\Omega_m=0.26$ $\Lambda$CDM).
These binned measurement are then translated from the ``fiducial'' cosmology to the measurements in a ``target'' cosmology using the following coordinate transforms,}
\begin{eqnarray}\label{eq:smumap}
 s_{\rm target} = s_{\rm fiducial} \sqrt{\alpha_{\parallel}^2 \mu_{\rm fiducial}^2+\alpha_{\bot}^2(1-\mu_{\rm fiducial}^2)}, \nonumber \\
 \mu_{\rm target} = \mu_{\rm fiducial} \frac{\alpha_\parallel}
 {\sqrt{\alpha_{\parallel}^2 \mu_{\rm fiducial}^2 +\alpha_{\bot}^2(1-\mu_{\rm fiducial}^2)}}
\end{eqnarray}
where $\alpha_{\bot}\equiv D_{A,\rm target}/D_{A,\rm fiducial}$,
$\alpha_{\parallel}\equiv H_{\rm fiducial}/H_{\rm target}$ 
and $D_A$ and $H$ are computed in the effective redshifts of the six redshift bins.
In the fiducial cosmology
we measure $\xi(s,\mu)$ with a high resolution of
$\Delta s = 0.2 {\rm Mpc/h}$, $\Delta \mu = 1/600$,
and later these small ``pixels'' are grouped to infer 
the number counts in other cosmologies, 
in large pixels of $\Delta s = 1 {\rm Mpc/h}$ and $\Delta \mu = 1/120$.
In the case when one small pixel belongs to more than one larger pixel,
a correction is applied by computing the fraction of the overlapping area
A dense grid of $\Delta s = 0.2 {\rm Mpc/h}$, $\Delta \mu = 1/600$ can significantly reduce the edge effect;
if we use Equation \ref{eq:smumap} to do a simple interpolation on a $\Delta s = 1 {\rm Mpc/h}$, $\Delta \mu = 1/120$  grid,
the edge effect becomes so large that the derived cosmological constraints suffer from a significant error.

\red{We tested the derived cosmological constraints from this approximation method compared to our \cite{Li2016} results, where we made the measurements in each cosmological model without approximation.  
Without considering the edge effect it deviates from the original contour by more than 1 sigma (even if the fiducial cosmology is the correct one).
Whatever the fiducial cosmology, the error always exist since we always need to compute $\chi^2$ of non-fiducial cosmological parameters when making the contour. The amplitude of error is found to be larger if the fiducial cosmology is far from the constrained region of parameter space. For example,
in the case that the deviation is as large as $\delta \Omega_m > 0.2$ and $\delta w > 2$, the change in the position and size of the contour is $\sim$10\%.
}

\begin{figure*}
   \centering{
   \includegraphics[width=12cm]{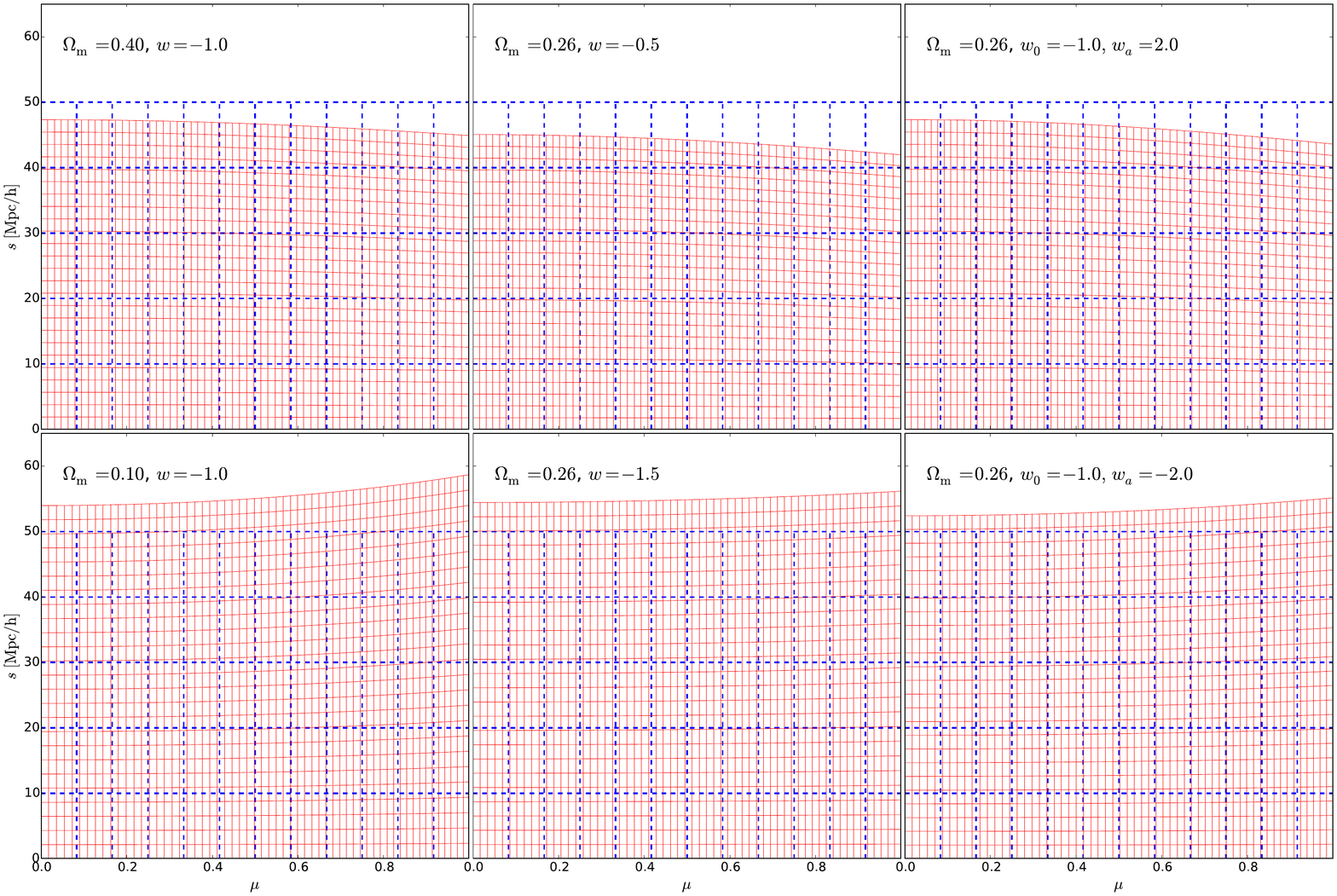}
   }
   \caption{\label{fig_method}
 Mapping $\xi(s,\mu)$ from the fiducial cosmology (taken as the $\Omega_m=0.26$ $\Lambda$CDM cosmology) to six different cosmologies,
   $(\Omega_m,w_0,w_a) = (0.4,-1,0),(0.26,-0.5,0), (0.26,-1,2), (0.1,-1,0), (0.26,-1.5,0),\ \rm and\ (0.26,-1,-2)$.
 The number counts are measured in the fiducial cosmology in the blue dashed grid;
 in other cosmologies their distribution becomes the red solid grid (according to Equation \ref{eq:smumap}).
 We use this relation to obtain $\xi(s,\mu)$ in these non-fiducial cosmology without re-measuring the number counts.
 To enhance the accuracy, we count the number of galaxy pairs in 5 times smaller pixels (the small red pixels), 
 and group them together to infer the values of $\xi(s, \mu)$ in the blue dashed pixels
 (for illustration purpose, the blue and red grids are 10 times sparser than the grids adopted in the real analysis).
}
\end{figure*}

The above procedure is illustrated in Figure \ref{fig_method}.
Using the relations given by Equation \ref{eq:smumap},
we obtained the distribution of number counts in cosmologies other than the fiducial cosmology.
We ensure the accuracy of the remapping by performing the pair counting using 5 times smaller pixels (the small red pixels), 
and regrouping these together to infer the number counts at the desired resolution (the large blue dashed pixels).

Figure \ref{fig_xi} plots the approximate 2PCFS in the $w_a=-2$, evolving dark energy, cosmology.
We find that, the approximation procedure only introduces a $\lesssim0.5\%$ error in the $\hat\xi_{\Delta s}(\mu)$, 
which is 10 times smaller than the intrinsic noise (the possion noise and cosmic variance) in $\hat\xi_{\Delta s}(\mu)$.
So it should be precise enough to use the approximate 2PCF in the statistical analysis.

\begin{figure*}
   \centering{
   \includegraphics[width=8.5cm]{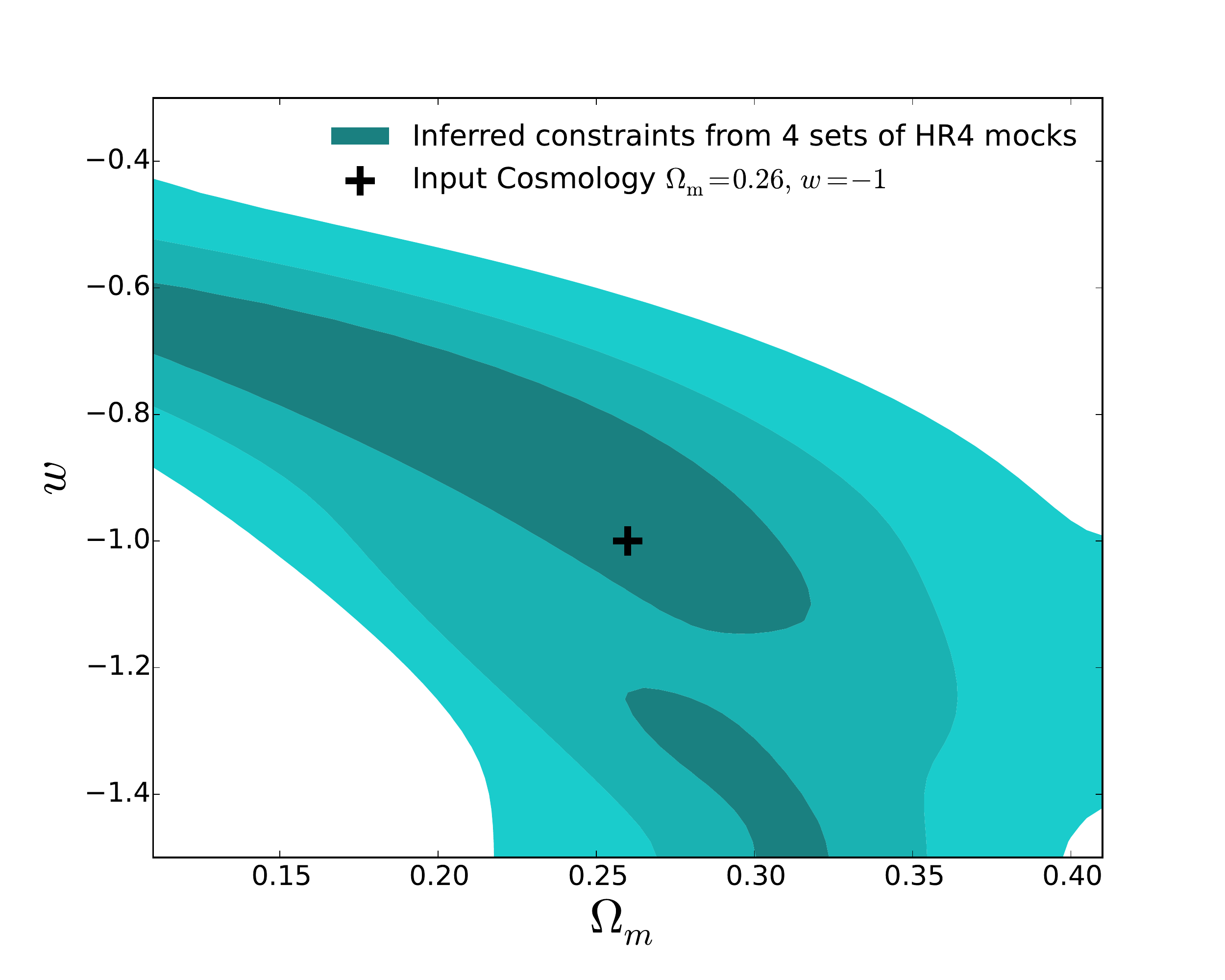}
   \includegraphics[width=8.5cm]{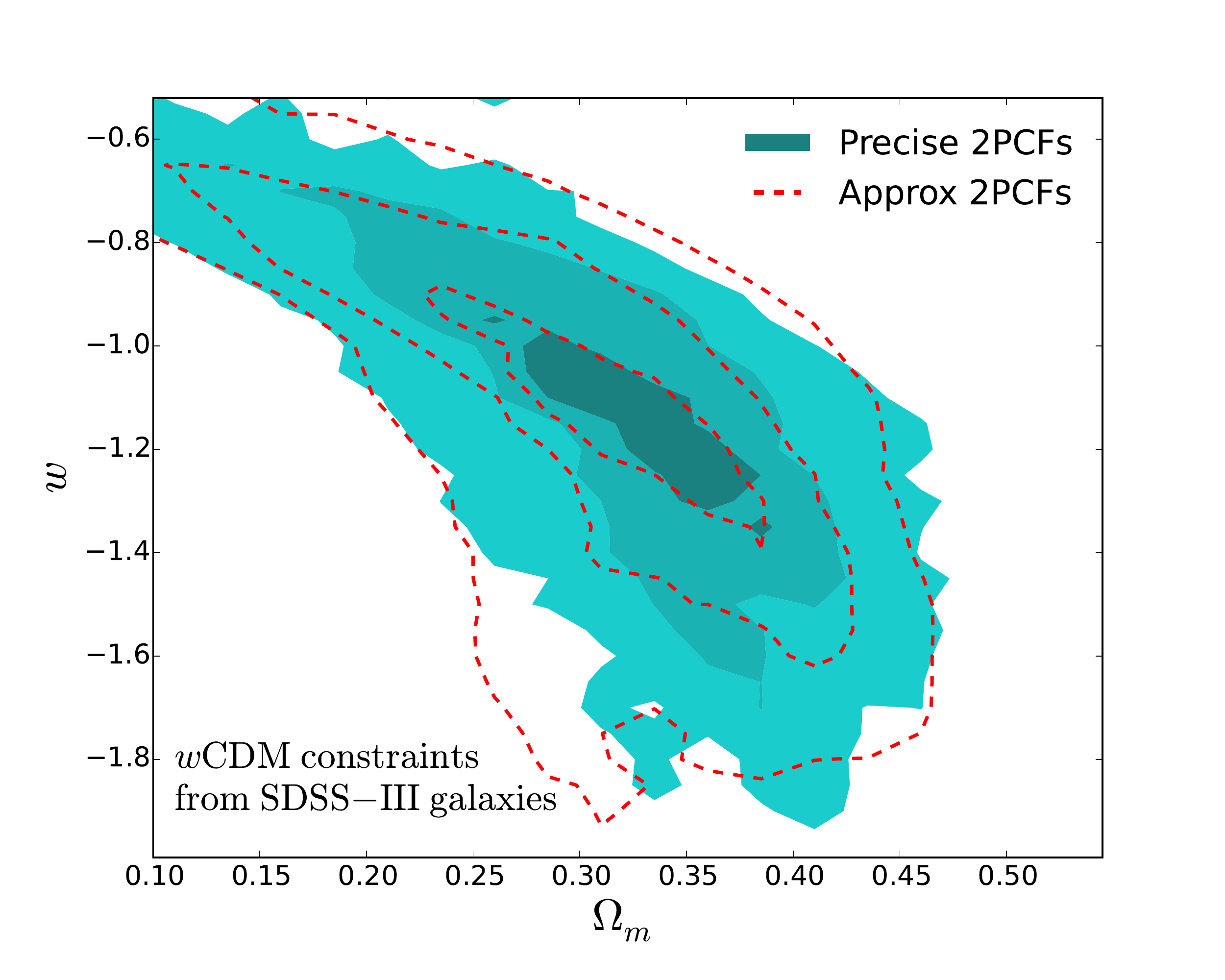}
   }
   \caption{\label{fig_Check}
 Left panel: Input-output test of the AP methodology using four sets of BOSS DR12 galaxy mock catalogues constructed from the HR4 simulation.
 The approximate 2PCFs are adopted in the analysis.
 The inferred cosmological constraints from the method, shown in the cyan contours, are consistent with the input cosmology
 (the simulation cosmology, i.e. the $\Omega_m=0.26$ $\Lambda$CDM; marked by the black plus sign).
 Right panel: Cosmological constraints from the BOSS DR12 galaxies, assuming a $w$CDM cosmology 
 (i.e., the value of $w_a$ fixed as zero).
 Results obtained using the precise 2PCFs (cyan filled) 
 and approximate 2PCFs (red dashed) agree with each other quite well.
   }
\end{figure*}

We performed a series of tests to check the reliability of using the approximate 2PCF in the cosmological analysis.
An input-output test was conducted using the four sets of mock catalogues of BOSS DR12 galaxies constructed from HR4 \citep{HR4}.
The results are shown in the left panel of Figure \ref{fig_Check}. 
The input cosmology is the simulation cosmology, i.e. $\Omega_m=0.26$ $\Lambda$CDM.
It lies within the 1$\sigma$ contour of the inferred constraints.
The right panel of Figure \ref{fig_Check} displays the cosmological constraints from real observational data
in case of fixing $w_a$ as 0 (hereafter $w$CDM).
Results obtained using the precise and approximate 2PCFs agree quite well with each other.

\section{Robustness Check}\label{sec:robust}
\begin{figure*}
   \centering{
   \includegraphics[width=\columnwidth]{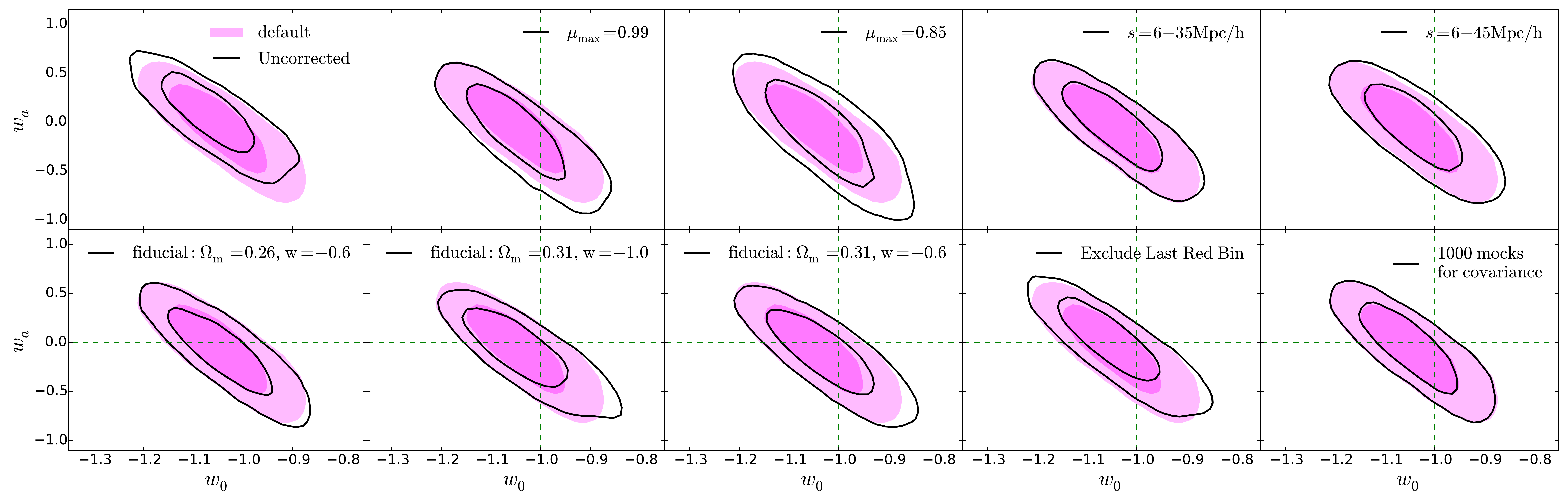}
   }
   \caption{\label{fig_contest}
   Robustness test of the results.
   The ``default'' constraints (magenta contours) are derived using six redshift bins, 
   $\mu_{\rm max}=0.97$, $s=6 - 40 \rm Mpc/h$,
   a fiducial cosmology $\Omega_m=0.26,\ w=-1.0$ for the approximation of 2PCF,
   systematic effects estimated from Horizon Run 4 simulations, 
   and covariance estimated using 2,000 MultiDark PATCHY mocks.
   When we alter one of these options by discarding the systematic correction, 
   using $\mu_{\rm max}=0.99$, 
   $\mu_{\rm max}=0.85$, $s=6 - 35 \rm Mpc/h$, $s=6 - 45 \rm Mpc/h$,  
   a fiducial cosmology of $\Omega_m=0.26,\ w=-0.6$, 
   excluding the last redshift bin in the analysis, 
   or reducing the number of mocks in the estimation of covariance matrix,
   the results remain robust (black contours).
   }
\end{figure*}

\begin{figure*}
   \centering{
   \includegraphics[width=0.6\columnwidth]{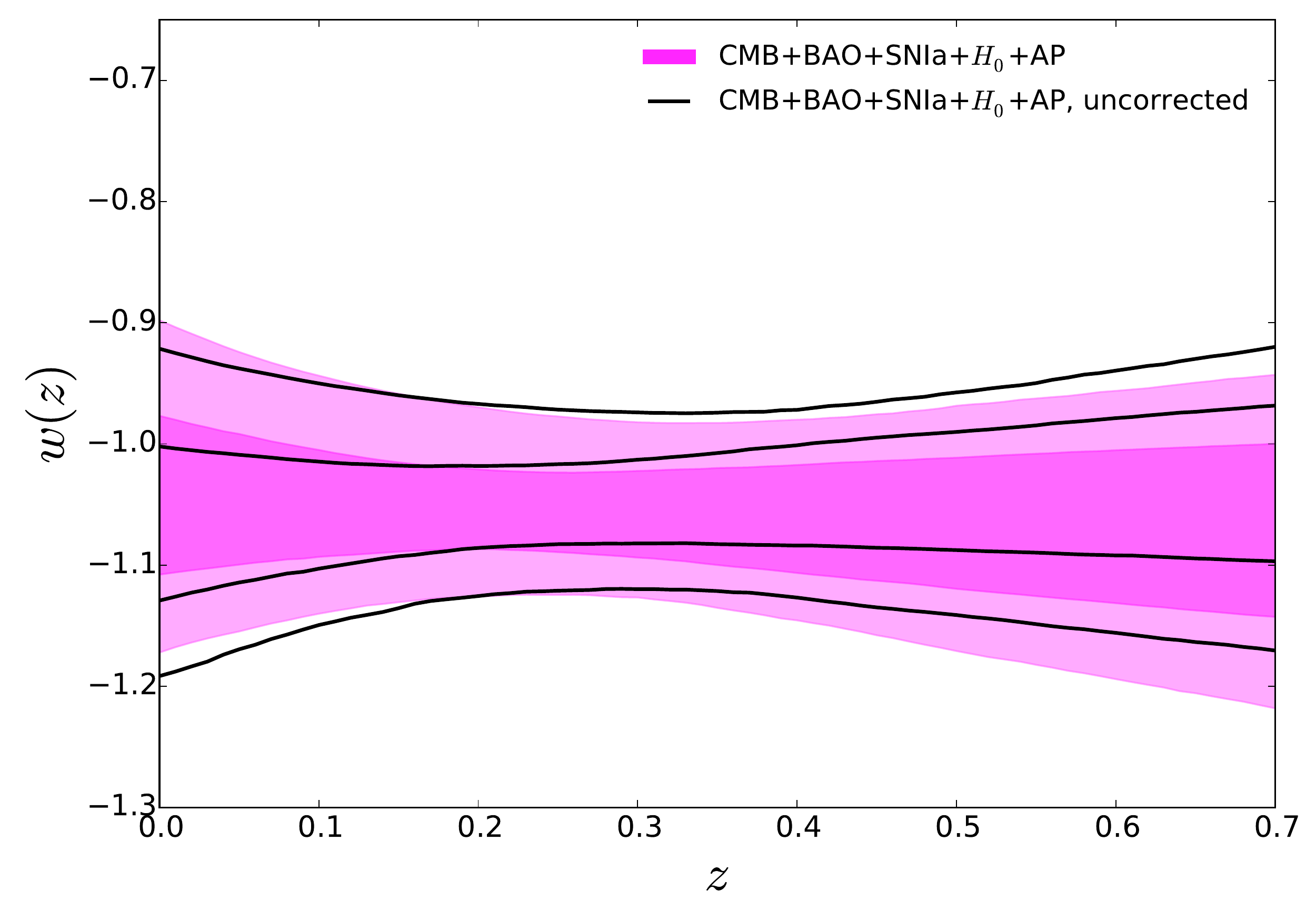}
   }
   \caption{\label{fig_wz_sys}
   Derived redshift evolution of $w(z)$ from CMB+BAO+SNIa+$H_0$+AP.
   There is no significant change if we discard the systematics correction procedure in the AP method analysis.
   }
\end{figure*}

Figure \ref{fig_contest} and \ref{fig_wz_sys} show that,
if we discard the systematics correction, 
the derived constraints are almost unaffected.
This indicates that, for the data analysis of current galaxy surveys,
the systematic effects in our method is not significant. 
But it remains to be seen if this is true for future galaxy surveys,
or when the cosmology dependence of the systematics effects is taken into account.

Furthermore, Figure \ref{fig_contest} shows that 
the result is unaffected by the line-of-sight $\mu$-cut, 
the range of radial integration, 
the choice of fiducial cosmology in the mapping of $\xi(s,\mu)$,
and the number of mocks.
The result does not change significantly if we remove the highest redshift bin,
where the estimated systematics is comparably large.
This further justifies our conclusion that, the effect of systematics is not significant in this analysis.

\end{document}